\documentclass[aps]{revtex4}
\usepackage{amsmath}
\usepackage{amssymb}
\usepackage{hyperref}
\usepackage{graphicx}
\usepackage{float}
\usepackage{slashed}
\hypersetup{colorlinks=true}
\begin{document}
\noindent
SLAC-PUB-17393

\title{Structure of light front vacuum sector diagrams}

\author{Philip D. Mannheim${}^1$, Peter Lowdon${}^2$ and  Stanley J. Brodsky${}^3$}
\affiliation{${}^1$ Department of Physics, University of Connecticut, Storrs, CT 06269, USA \\
${}^2$ Centre de Physique Th\'{e}orique, \'{E}cole polytechnique, F-91128 Palaiseau, France\\
${}^3$ SLAC National Accelerator Laboratory, Stanford University, Menlo Park, CA 94025, USA\\
philip.mannheim@uconn.edu,~peter.lowdon@polytechnique.edu,~sjbth@slac.stanford.edu}

\date{April 10, 2019}

\begin{abstract}
We study the structure of scalar field light front quantization vacuum graphs. In instant time quantization both non-vacuum and vacuum graphs can equivalently be described by either the off-shell four-dimensional Feynman diagram approach or the on-shell three-dimensional Fock space approach, with this being the case since the relevant Feynman diagrams are given entirely by pole terms. This is also the case for light front quantization non-vacuum graphs. However this is the not the case for light front vacuum sector diagrams, since then there are also circle at infinity contributions to Feynman diagrams. These non-pole contributions cause vacuum diagrams to be nonzero and to not be given by a light-front Hamiltonian Fock space analysis. The three-dimensional approach thus fails in the light front vacuum sector. In consequence, the closely related infinite momentum frame approach also fails in the light front vacuum sector. 

 \end{abstract}

\maketitle

\section{Introduction}
\label{S1}

Since the original work of Dirac \cite{Dirac:1949cp}, there has been a continuing interest in light front (also known as  ``light cone' or ``front form') quantization of quantum field theories.  Comprehensive reviews can be found in \cite{Brodsky:1997de,Leutwyler:1977vy,Bakker:2013cea,Burkardt:1995ct}.
The light-front approach is  based on 3-dimensional Hamiltonian field theory quantized at fixed light front time $ x^+= x^0+ x^3$.  The rules for calculations for Light-Front Hamiltonian QCD for both perturbative and nonperturbative applications are summarized in \cite{Lepage:1980fj}.  As is the case with the standard four-dimensional covariant Feynman Lagrangian theory,  the light-front formalism is Poincar\'e invariant and causal.  Observables in hadron physics such as form factors, structure functions, and distribution amplitudes are based on the nonperturbative light-front hadronic wavefunctions, the eigenfunctions of the QCD Light Front Hamiltonian~\cite{Lepage:1979zb,Brodsky:2003pw}.  In the case of scattering amplitudes, the covariant Feynman and the Light Front Hamiltonian approaches give identical results.  One can also replicate the calculation rules for light front $x^+$-ordered perturbation theory using standard time-ordered perturbation theory based on quantization at fixed time (also known as  instant time or ``instant form")  by choosing a Lorentz frame where the observer moves at infinite momentum \cite{Weinberg:1966jm,Brodsky:1973kb,Chang1969,Yan1973}. 
 
While the light front non-vacuum (i.e., scattering) sector is well understood, in the light front literature  there has been a spirited discussion as to the status of perturbative light front vacuum graphs (see e.g. \cite{Chang1969,Yan1973,Brodsky:1997de,Casher1974,Brodsky:2009zd,Collins2018}). In the light front vacuum sector differing results have been obtained for  the off-shell four-dimensional Feynman diagram approach and the on-shell three-dimensional  Fock space approach, and  the literature has not yet settled on which particular one might have fundamental validity, or identified what it is that causes differences between the various approaches. It is the purpose of this paper to address this issue in the scalar field theory case, and to show that because of circle at infinity contributions in four-dimensional light front vacuum Feynman diagrams it is the Feynman approach that one must use as the light front Fock space approach is equivalent to the pole term contribution to Feynman diagrams alone. Because of these non-pole circle at infinity contributions, light front vacuum diagrams are not only nonzero, they are equal to instant time vacuum diagrams, even though instant time vacuum Feynman  diagrams receive no circle at infinity contributions. Our result is initially surprising since the instant time Fock space analysis correctly describes the instant time vacuum sector, and in the infinite momentum frame the instant time Fock space procedure transforms into the light front Fock space description. However, even though circle at infinity contributions are suppressed in the instant time case, when instant time vacuum graphs are evaluated in the infinite momentum frame we find that instant time circle contributions are no longer suppressed, to thus cause the light front Fock space procedure to fail in the light front vacuum sector. Thus one must use the off-shell four-dimensional Feynman diagram approach in order to correctly describe the light front vacuum sector. Since circle at infinity issues are not of relevance in either  instant time non-vacuum or instant time vacuum graphs, and since circle at infinity issues are not even of relevance in the light front non-vacuum sector, we see that the light front vacuum sector has an intrinsic structure that is all its own, and it has to be treated independently.

\begin{figure}[H]
\begin{center}
\includegraphics[scale=0.15]{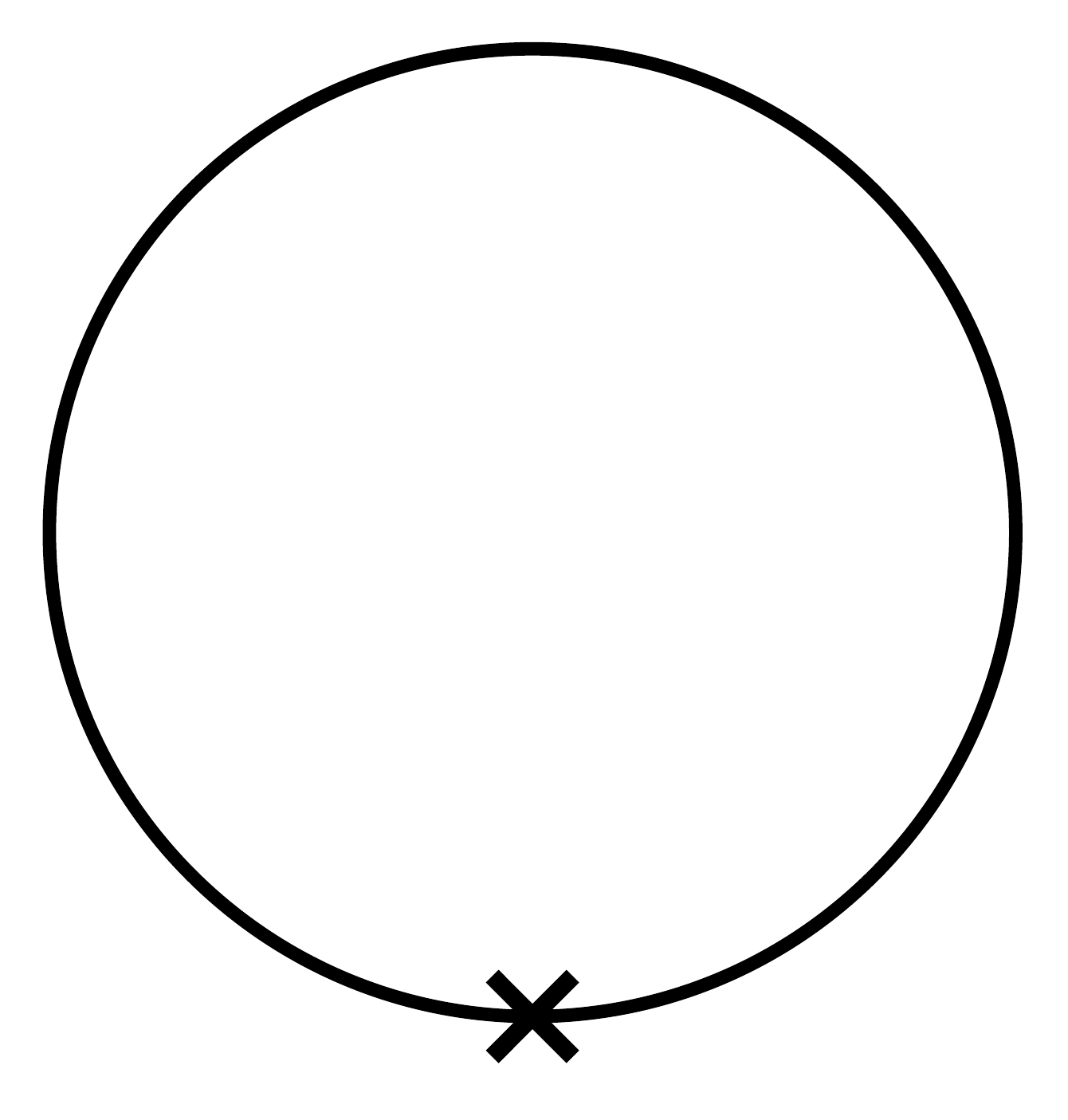}
\caption{Connected $\langle\Omega|\phi(0)\phi(0)|\Omega\rangle$}
\label{vacuumtadpole}
\end{center}
\end{figure}

To address these issues we have found it instructive to study Feynman diagrams in coordinate space rather than in momentum space, and in order to establish our results we only need to study the free propagators that contribute in a perturbative expansion, as they will prove rich enough for our purposes here. We shall thus study the structure of the scalar field $D(x^{\mu})=-i\langle \Omega|T[\phi(x)\phi(0)]|\Omega\rangle$ and its $x^{\mu}\rightarrow 0$ limit $D(x^{\mu}=0)=-i\langle \Omega|\phi(0)\phi(0)|\Omega\rangle$, and this will enable us to use the spacetime coordinates as regulators when we  take the limit.  For the  action $I_S=\int d^4x(-g)^{1/2}[\tfrac{1}{2}\partial_{\mu}\phi\partial^{\mu}\phi-\tfrac{1}{2}m^2\phi^2]$, the free $D(x^{\mu})$ propagator is given as a Feynman diagram as:
\begin{eqnarray}
D(x^{\mu})=-i\langle \Omega|[\theta(\sigma)\phi(x)\phi(0)+\theta(-\sigma)\phi(0)\phi(x)]|\Omega\rangle=\frac{1}{(2\pi)^4}\int d^4p \frac{e^{-ip\cdot x}}{p^2-m^2+i\epsilon}.
\label{FV1}
\end{eqnarray}
where $\sigma$ is $x^0$ in the instant time case and is $x^+=x^0+x^3$ in the light front case. With $\theta(0)=1/2$ the associated $x^{\mu}=0$ vacuum graph is given by 
\begin{eqnarray}
D(x^{\mu}=0)=-i\langle \Omega|\phi(0)\phi(0)|\Omega\rangle=\frac{1}{(2\pi)^4}\int d^4p \frac{1}{p^2-m^2+i\epsilon}.
\label{FV2}
\end{eqnarray}
and is represented by the graph shown in Fig. \ref{vacuumtadpole}. The graph can be represented as a circle with a cross on the circumference, with the cross representing a $\phi^2$ insertion. (Without the cross the graph would represent a disconnected graph and would not be of interest.) The graph shown in Fig. \ref{vacuumtadpole} would occur as a connected one loop tadpole graph in a $\lambda \phi^3$ theory, with an amputated external $\phi$ field bringing zero momentum into the cross where two $\phi$ fields are created with strength $\lambda$. It is thus the limit in which coordinate space points are brought together. For our purposes here it will prove to be more instructive to treat the tadpole graph as the $x^{\mu}\rightarrow 0$ limit of a Feynman time-ordered propagator rather than the  limit $x^{\mu}\rightarrow 0$ of a two-point function such as $-i\langle \Omega |\phi(x)\phi(0)|\Omega\rangle$.  The tadpole graph would also appear in a $g\phi\bar{\psi}\psi$ theory with the cross representing a fermion-antifermion insertion at the point where the scalar $\phi$ brings zero momentum into the loop with strength $g$. The tadpole graph appears in mass renormalization in theories such as $\lambda\phi^4$ or $g(\bar{\psi}\psi)^2$, in theories of dynamical symmetry breaking  by fermion or scalar field bilinear condensates, and in gravity theories where it can couple to the trace of  the matter energy-momentum tensor, and thus be of relevance to cosmology and the cosmological constant problem. 

When written out in detail we have
\begin{eqnarray}
D(x^{\mu},{\rm instant})&=&\frac{1}{(2\pi)^4}\int dp_0dp_1dp_2dp_3 \frac{e^{-i(p_0x^0+p_1x^1+p_2x^2+p_3x^3)}}{(p_0)^2-(p_1)^2-(p_2)^2-(p_3)^2-m^2+i\epsilon},
\nonumber\\
D(x^{\mu},{\rm front})&=&\frac{2}{(2\pi)^4}\int dp_+dp_1dp_2dp_- \frac{e^{-i(p_+x^++p_1x^1+p_2x^2+p_-x^-)}}{4p_+p_--(p_1)^2-(p_2)^2-m^2+i\epsilon},
\nonumber\\
D(x^{\mu}=0,{\rm instant})&=&\frac{1}{(2\pi)^4}\int dp_0dp_1dp_2dp_3 \frac{1}{(p_0)^2-(p_1)^2-(p_2)^2-(p_3)^2-m^2+i\epsilon},
\nonumber\\
D(x^{\mu}=0,{\rm front})&=&\frac{2}{(2\pi)^4}\int dp_+dp_1dp_2dp_- \frac{1}{4p_+p_--(p_1)^2-(p_2)^2-m^2+i\epsilon}.
\label{FV3}
\end{eqnarray}
where $x^-=x^0-x^3$ \cite{footnote1}. As long as either $x^0$ or $x^+$ is nonzero and positive, the circle at infinity contributions in the lower half of the complex $p_0$ or $p_+$ planes are suppressed by the $e^{-ip\cdot x}$ term, and the only contributions to the Feynman contours are the pole terms. Similarly, even if $x^{0}$ is zero, the circle at infinity contribution to the instant $D(x^{\mu}=0,{\rm instant})$ is still suppressed because there are two powers of $p_0$ in the denominator. However, the circle at infinity contribution to the front $D(x^{\mu}=0,{\rm front})$ is not suppressed because in that case there is only one power of $p_-$ in the denominator. It is in this way then that $D(x^{\mu}=0,{\rm front})$ is conceptually different.

This outcome is initially somewhat surprising since the $\theta$ functions (technically distributions) in the definition of time ordered products can be written as contour integrals of the form
\begin{eqnarray}
\theta(\sigma)=-\frac{1}{2\pi i}\oint_{-\infty}^{\infty} d\omega \frac{e^{-i\omega \sigma}}{\omega+i\epsilon},
\label{FV4}
\end{eqnarray}
(to thus cause Feynman diagrams to be off shell). And our ability to show that the complex $\omega$ plane contour integral is in fact a $\theta$ function resides in the fact that because of the $e^{-i\omega \sigma}$ term the circle at infinity contribution along a contour in the lower half complex $\omega$ plane is suppressed if $\sigma>0$, with the pole term giving $-(1/2\pi i)\times (-2\pi i)=1$ if $\sigma>0$, However, suppose we drop the suppression factor by setting $\sigma=0$. We now have a circle at infinity contribution and obtain 
\begin{eqnarray}
\theta(0)=-\frac{1}{2\pi i}\oint_{-\infty}^{\infty} d\omega \frac{1}{\omega+i\epsilon}=-\frac{1}{2\pi i}[
-2\pi i+\pi i]=\frac{1}{2}
\label{FV5}
\end{eqnarray}
just as we should (i.e. our particular representation of $\theta(\sigma)$ as the contour integral given in (\ref{FV3}) entails that $\theta(0)=1/2$), and just as required in going from (\ref{FV1}) to (\ref{FV2}). Thus if we construct $D(x^{\mu}=0,{\rm front})$ as the $x^{\mu}=0$ limit of $D(x^{\mu},{\rm front})$, then the circle at infinity term that had been suppressed when $x^{\mu}\neq 0$ is no longer suppressed, and thus needs to be taken into consideration.

\section{The Non-Vacuum instant time Case}

In the instant time case the Feynman integral is readily performed since it is just pole terms and yields 
\begin{eqnarray}
D(x^0>0,{\rm instant})&=&D(x^0>0,{\rm instant},{\rm pole})
\nonumber\\
&=&-\frac{i}{(2\pi)^3}\int_{-\infty}^{\infty} \frac{d^3p}{2E_p}\exp(-iE_px^0+i\vec{p}\cdot \vec{x})
=\frac{1}{8\pi}\left(\frac{m^2}{x^2}\right)^{1/2}H^{(2)}_1(m(x^2)^{1/2}),
\label{FV6}
\end{eqnarray}
where $E_p=+(\vec{p}^2+m^2)^{1/2}$. In the instant time case one can take an instant time forward Green's function such as $D(x^0>0,{\rm instant})=-i\langle \Omega_I|\theta(x^0)\phi(x^0,x^1,x^2,x^3)\phi(0)|\Omega_I\rangle$ as evaluated in the instant time no-particle (viz. vacuum) state $|\Omega_I\rangle$, and expand the field in terms of instant time creation and annihilation operators that create and annihilate particles out of that vacuum state as  
\begin{eqnarray}
\phi(\vec{x},x^0)=\int \frac{d^3p}{(2\pi)^{3/2}(2E_p)^{1/2}}[a(\vec{p})\exp(-iE_p t+i\vec{p}\cdot\vec{x})+a^{\dagger}(\vec{p})\exp(+iE_p t-i\vec{p}\cdot\vec{x})],
\label{FV7}
\end{eqnarray}
where $[a(\vec{p}),a^{\dagger}(\vec{p}^{\prime})]=\delta^3(\vec{p}-\vec{p}^{\prime})$. The insertion of $\phi(\vec{x},x^0)$ into $D(x^0>0,{\rm instant})$ immediately leads to the on-shell three-dimensional integral
\begin{eqnarray}
D(x^0>0,{\rm instant,~Fock})=-\frac{i\theta(x^0)}{(2\pi)^3}\int_{-\infty}^{\infty} \frac{d^3p}{2E_p} e^{-iE_p x^0+i\vec{p}\cdot\vec{x}}.
\label{FV8}
\end{eqnarray}
We recognize (\ref{FV8}) as (\ref{FV6}), to thus establish the equivalence of the instant time Feynman and Fock space prescriptions.

\section{The Non-Vacuum light front Case}

In the light front case poles in the complex $p_+$ plane occur at
\begin{eqnarray}
p_+=E^{\prime}_p-\frac{i\epsilon}{4p_-},
\label{FV9}
\end{eqnarray}
where $E^{\prime}_p=((p_1)^2+(p_2)^2+m^2)/4p_-$. Poles with $p_-\geq  0^+$ thus all lie below the real $p_+$ axis and have positive $E_p^{\prime}$, while poles with $p_-\leq 0^-$ all lie above the real $p_+$ axis and have negative $E_p^{\prime}$. For $x^+>0$, closing the $p_+$ contour below the real axis then restricts the poles to $E^{\prime}_p>0$, $p_-\geq 0^+$. However, in order to evaluate the pole terms one has to deal with the fact that the pole at $p_-=0^+$ has $E_p^{\prime}=\infty$. We shall thus momentarily exclude the region around $p_-=0$, and thus only consider poles below the real $p_+$ axis that have $p_-\geq \delta$ where $\delta$ is a small positive number. Evaluating the contour integral in the lower half of the complex $p_+$ plane thus gives
\begin{align}
D(x^+>0,{\rm front},{\rm pole})&=-\frac{2i}{(2\pi)^3}\int_{\delta}^{\infty}\frac{dp_-}{4p_-}\int_{-\infty}^{\infty} dp_1\int_{-\infty}^{\infty}dp_2 
e^{-i(E^{\prime}_px^++p_-x^-+p_1x^1+p_2x^2)-\epsilon x^+/4p_-}
\nonumber\\
&=-\frac{1}{4\pi^2x^+}\int_{\delta}^{\infty} dp_-e^{-ip_-x^-+i[(x^1)^2+(x^2)^2]p_-/x^+-im^2x^+/4p_--\epsilon x^+/4p_-}
\nonumber\\
&=-\frac{1}{4\pi^2x^+}\int_{\delta}^{\infty}dp_-e^{-ip_-x^2/x^+-im^2x^+/4p_--\epsilon x^+/4p_-}.
\label{FV10}
\end{align}
If we now set $\alpha=x^+/4p_-$, we obtain
\begin{align}
D(x^+>0,{\rm front},{\rm pole})=-\frac{1}{16\pi^2}\int_{0}^{x^+/4\delta} \frac{d\alpha}{\alpha^2}e^{-ix^2/4\alpha-i\alpha m^2-\alpha\epsilon}.
\label{FV11}
\end{align}
In (\ref{FV11}) we can now take the limit $\delta \rightarrow 0$, $x^+/4\delta\rightarrow \infty$ without encountering any ambiguity as long as $x^+$ is nonzero, and with $x^+>0$ thus obtain
\begin{align}
D(x^+>0,{\rm front},{\rm pole})=-\frac{1}{16\pi^2}\int_{0}^{\infty} \frac{d\alpha}{\alpha^2}e^{-ix^2/4\alpha-i\alpha m^2-\alpha\epsilon}.
\label{FV12}
\end{align}
This integral is readily done and yields 
\begin{align}
D(x^+>0,{\rm front})=D(x^+>0,{\rm front},{\rm pole})=\frac{1}{8\pi}\left(\frac{m^2}{x^2}\right)^{1/2}H^{(2)}_1(m(x^2)^{1/2}).
\label{FV13}
\end{align}

Comparing with (\ref{FV6}) we see that $D(x^+>0,{\rm instant})$ and $D(x^+>0,{\rm front})$ are equal. As discussed in \cite{Mannheim2019}, where details of our work may be found,  this is not the case just for this particular Green's function, as it actually holds  for the instant time and light front evaluations of any scalar field Green's function.  Specifically, while one ordinarily tries to relate instant time and light front graphs by a Lorentz boost to the infinite momentum frame (something we discuss below), the transformation $x^0\rightarrow x^+=x^0+x^3$, $x^3\rightarrow x^-=x^0-x^3$ is actually a spacetime dependent translation, i.e. a general coordinate transformation. Since Feynman diagrams are just integrals over c-number momentum variables, and when written in coordinate space are just functions of c-number coordinates, Feynman diagrams are general coordinate invariant, and thus instant time and light front evaluations of any scalar field Feynman diagram must be equal \cite{footnote1a}. However, that does not mean that one can transform equal instant time canonical commutators into equal front time ones as well. And in fact one cannot \cite{Mannheim2019}, and as noted in \cite{Mannheim2019} it is this that enables the instant time and light front scalar field Green's functions to be equal.

\section{Non-Vacuum light front Fock Space Treatment}

In the light front case the Fock space expansion of modes that obey $[4\partial_-\partial_+-(\partial_1)^2-(\partial_2)^2+m^2]\phi(x^+,x^1,x^2,x^-)=0$ is of the form 
\begin{align}
\phi(x^+,x^-,x^1,x^2)&=\frac{2}{(2\pi)^{3/2}}\int_{-\infty}^{\infty}dp_1\int_{-\infty}^{\infty}dp_2\int_0^{\infty} \frac{dp_-}{(4p_-)^{1/2}}
\bigg{[}e^{-i(F^2_px^+/4p_-+p_-x^-+p_1x^1+p_2x^2)}a_p
\nonumber\\
&+e^{i(F^2_px^+/4p_-
+p_-x^-+p_1x^1+p_2x^2)}a_p^{\dagger}\bigg{]},
\label{FV14}
\end{align}
where $F^2_p=(p_1)^2+(p_2)^2+m^2$, and where the light front $[a_p,a^{\dagger}_{p^{\prime}}]$ commutator  is normalized to $[a_p,a^{\dagger}_{p^{\prime}}]=(1/2)\delta(p_--p_-^{\prime})\delta(p_1-p_1^{\prime})\delta(p_2-p_2^{\prime})$ so as to impose the canonical commutator $[\phi(x^+,x^1,x^2,x^-),2\partial_-\phi(x^+,y^1,y^2,y^-)]=i\delta(x^1-y^1)\delta(x^2-y^2)\delta(x^--y^-)$, a derivation for which may be found in \cite{Neville1971} and more recently in \cite{Mannheim2019}. In (\ref{FV14}) we note that the $p_-$ integration is only over nonnegative $p_-$ \cite{footnote2}.

With this normalization we can then insert this on-shell form for $\phi(x)$ into $D(x^{\mu},{\rm front})=-i\langle \Omega_F |[\theta(x^+)\phi(x^{\mu})\phi(0)+\theta(-x^+)\phi(0)\phi(x^{\mu})]|\Omega_F\rangle$ as evaluated in the light front no-particle state $|\Omega_F\rangle$ that the light front $a_p$ annihilate. With the insertion of this $\phi(x)$ then precisely giving the first line in (\ref{FV10}), we establish the equivalence of the Feynman and Fock space prescriptions in the non-vacuum light front case. As we see, just as with the instant time case, since there are only pole terms and no circle at infinity contributions  in the non-vacuum case, we are able to establish the equivalence of the light front Fock space and Feynman diagram prescriptions in such case. In the non-vacuum (i.e. $x^+\neq 0$) sector we  are thus able to validate the standard non-vacuum light front on-shell Fock space prescription that is widely used in light front studies. And in addition we see a general rule emerge, namely that  Feynman and Fock space prescriptions will coincide whenever the only contribution to Feynman contours is poles. However, as we shall now see, in the light front vacuum sector there are circle at infinity contributions, to thus cause the on-shell Fock space description to become invalid.

\section{The instant time Vacuum Case}

In the instant time case one can readily set $x^{\mu}$ to zero in (\ref{FV6}), (\ref{FV7}) and (\ref{FV8}), to obtain
\begin{align}
D(x^{\mu}=0,{\rm instant})&=D(x^{\mu}=0,{\rm instant},{\rm pole})=D(x^{\mu}=0,{\rm instant},{\rm Fock})
\nonumber\\
&
=-\frac{i}{(2\pi)^3}\int_{-\infty}^{\infty} \frac{d^3p}{2E_p}
=-\frac{1}{16\pi^2}\int_0^{\infty}\frac{d\alpha}{\alpha^2}e^{-i\alpha m^2-\alpha\epsilon}.
\label{FV15}
\end{align}

\section{The light front Vacuum Case -- Pole  Contribution}

In the light front case we set $x^{\mu}$ to zero and evaluate $D(x^{\mu}=0,{\rm front})$ as given in (\ref{FV3}). Just as  above we again need to take care of the $p_-=0$ region, so we again introduce the $\delta$ cutoff at small $p_-$. On closing below the real $p_+$ axis the only poles are those with $p_->0$, and for them we obtain a pole contribution of the form
\begin{align}
D(x^{\mu}=0,{\rm front},{\rm pole})&=-\frac{2i}{(2\pi)^3}\int_{-\infty}^{\infty} dp_1\int_{-\infty}^{\infty}dp_2 \int_{\delta}^{\infty}\frac{dp_-}{4p_-}.
\label{FV16}
\end{align}
Then on setting $p_-=1/\alpha$, we are able to let $p_-$ go to zero,  to obtain 
\begin{align}
D(x^{\mu}=0,{\rm front},{\rm pole})&=-\frac{i}{16\pi^3}\int_{-\infty}^{\infty} dp_1\int_{-\infty}^{\infty}dp_2 \int_{0}^{1/\delta}\frac{d\alpha}{\alpha}
=-\frac{i}{16\pi^3}\int_{-\infty}^{\infty} dp_1\int_{-\infty}^{\infty}dp_2 \int_{0}^{\infty}\frac{d\alpha}{\alpha}.
\label{FV17}
\end{align}
For the Fock space prescription we set $x^{\mu}=0$ in (\ref{FV14}), viz.
\begin{align}
\phi(0)&=\frac{2}{(2\pi)^{3/2}}\int_{-\infty}^{\infty}dp_1\int_{-\infty}^{\infty}dp_2\int_0^{\infty} \frac{dp_-}{(4p_-)^{1/2}}
[a_p+a_p^{\dagger}],
\label{FV18}
\end{align}
and on inserting $\phi(0)$  into $-i\langle \Omega|\phi(0)\phi(0)]|\Omega\rangle$ obtain 
\begin{align}
D(x^{\mu}=0,{\rm front},{\rm Fock})=-\frac{2i}{(2\pi)^3}\int_{-\infty}^{\infty} dp_1\int_{-\infty}^{\infty}dp_2 \int_{0}^{\infty}\frac{dp_-}{4p_-}.
\label{FV19}
\end{align}
Comparing with (\ref{FV16}) we again see the equivalence of the pole and Fock space prescriptions. 

However, there is something wrong with both prescriptions. We are evaluating the $m$-dependent $D(x^{\mu}=0,{\rm front})$ as given in (\ref{FV3}), and yet we obtain an answer that does not depend on $m$ at all. Moreover, the theorem presented in \cite{Mannheim2019} and \cite{footnote1a} would require that $D(x^{\mu}=0,{\rm front})$ and $D(x^{\mu}=0,{\rm instant})$ be equal, and neither $D(x^{\mu}=0,{\rm front},{\rm pole})$ nor $D(x^{\mu}=0,{\rm front},{\rm Fock})$ is equal to $D(x^{\mu}=0,{\rm instant})$ as given in (\ref{FV15}), and indeed they could not be since $D(x^{\mu}=0,{\rm instant})$ is $m$-dependent. Thus something must have gone wrong.

\section{The light front Vacuum Case -- Circle at Infinity  Contribution}

What went wrong is that there is a circle at infinity contribution. To evaluate the circle at infinity contribution we have found it convenient to use an exponential regulator on the circle of the type usually used on the real frequency axis as this enables us to set $1/(A+i\epsilon)=-i\int_0^{\infty}\exp(i\alpha (A+i\epsilon))$ for any $A+i\epsilon$ on the circle that is such that there is convergence at $\alpha=\infty$. On setting 
\begin{align}
D(x^{\mu}=0,{\rm front},{\rm circle})=-\frac{2i}{(2\pi)^4}\int_{-\infty}^{\infty} dp_+\int_{-\infty}^{\infty}dp_-\int_{-\infty}^{\infty}dp_1\int_{-\infty}^{\infty}dp_2\int_0^{\infty}
d\alpha e^{i\alpha(4p_+p_--(p_1)^2-(p_2)^2-m^2+i\epsilon)},
\label{FV20}
\end{align}
we see that on setting $p_+=Re^{i\theta}$ on a circle at infinity of radius $R$ we can get convergence at $\alpha=\infty$ if $4i\alpha p_-R(\cos\theta+i\sin\theta)=4i\alpha p_-R\cos\theta-4\alpha p_-R\sin\theta$ converges, i.e. if $p_-\sin\theta$ is positive. With positive $p_-$ this would then require that $\sin\theta$ be positive, while negative $p_-$ would require that $\sin\theta$ be negative. Now $\sin\theta$ is positive for $0<\theta <\pi$, and negative for $\pi<\theta<2\pi$. Thus in order to use the exponential regulator on the circle we must close above the real $p_+$ axis for positive $p_-$, while  we must close below the real $p_+$ axis for negative $p_-$. However, for positive $p_-$ the poles in $p_+$ are below the real axis, while for negative $p_-$ the poles in $p_+$ are above the real axis. Thus in applying the exponential regulator on the circle at infinity we always have to close the contour so that we do not encounter any poles at all, to thereby show that the circle contribution cannot be ignored.  As we see from (\ref{FV20}), the great utility of the use of the exponential regulator in the light front case is that it is well-defined at $p_-=0$. 

Symbolically we can  set 
\begin{eqnarray}
\int_{-\infty}^{\infty} dp_+=\int_{-\infty}^{\infty} dp_+(p_->0)+\int_{-\infty}^{\infty} dp_+(p_-<0)=-\int_{0}^{\pi}d\theta(p_->0)-\int_{2\pi}^{\pi}d\theta(p_-<0).
\label{FV21}
\end{eqnarray}
And thus for $p_->0$ first we obtain an upper circle contribution to $D(x^{\mu}=0,{\rm front})$ of the form
\begin{align}
&D(x^{\mu}=0,p_->0,{\rm front},{\rm upper~circle})
\nonumber\\
&=\frac{2i}{(2\pi)^4}\int_0^{\infty}dp_-\int_{-\infty}^{\infty} dp_1\int_{-\infty}^{\infty}dp_2\int _0^{\pi} iRe^{i\theta}d\theta\int_0^{\infty}d\alpha e^{i\alpha(4p_-Re^{i\theta}-(p_1)^2-(p_2)^2-m^2+i\epsilon)}
\nonumber\\
&=\frac{1}{8\pi^3}\int_0^{\infty}dp_-\int_0^{\infty}\frac{d\alpha}{\alpha}e^{-i\alpha m^2-\alpha \epsilon}
\int _0^{\pi} iRe^{i\theta}d\theta e^{4i\alpha p_-Re^{i\theta}}
\nonumber\\
&=\frac{1}{8\pi^3}\int_0^{\infty}dp_-\int_0^{\infty}\frac{d\alpha}{\alpha}e^{-i\alpha m^2-\alpha \epsilon}\frac{(e^{-4i\alpha p_-R}-e^{4i\alpha p_-R})}{4i\alpha p_-}
\nonumber\\
&=-\frac{1}{4\pi^3}\int_0^{\infty}dp_-\int_0^{\infty}\frac{d\alpha}{\alpha}e^{-i\alpha m^2-\alpha \epsilon}
\frac{\sin(4\alpha p_-R)}{4\alpha p_-}.  
\label{FV22}
\end{align}
Then, on letting $R$ go to infinity we obtain
\begin{align}
&D(x^{\mu}=0,p_->0,{\rm front},{\rm upper~circle})=-\frac{1}{4\pi^2}\int_0^{\infty}dp_-\int_0^{\infty}\frac{d\alpha}{\alpha}e^{-i\alpha m^2-\alpha \epsilon}
\delta(4\alpha p_-) 
\nonumber\\
&=-\frac{1}{8\pi^2}\int_{-\infty}^{\infty}dp_-\int_{0}^{\infty}\frac{d\alpha}{\alpha}e^{-i\alpha m^2-\alpha \epsilon}\delta(4\alpha p_-) 
=-\frac{1}{32\pi^2}\int_{0}^{\infty}\frac{d\alpha}{\alpha^2}e^{-i\alpha m^2-\alpha \epsilon}.
\label{FV23}
\end{align}
As the presence of the $\delta(4\alpha p_-)$ term shows, the key region is $p_-=0$, something also noted in \cite{Chang1969,Yan1973}. In an on-shell approach states obey $4p_+p_--(p_1)^2-(p_2)^2=m^2$, and thus states with $p_-=0$ would be missed, and without them one would otherwise have had to conclude \cite{Brodsky:1997de} that $D(x^{\mu}=0,{\rm front})=0$ \cite{footnote3}. 

Moreover, since $\sin 0=\sin 2\pi$ and since $\delta(4\alpha p_-)$ is even under $p_-\rightarrow -p_-$, it follows that $D(x^{\mu}=0,p_->0,{\rm front},{\rm upper~circle})$ and $D(x^{\mu}=0,p_-<0,{\rm front},{\rm lower~circle})$ must be equal. Thus finally we obtain
\begin{align}
D(x^{\mu}=0,{\rm front})&=D(x^{\mu}=0,p_->0,{\rm front},{\rm upper~circle})+D(x^{\mu}=0,p_-<0,{\rm front},{\rm lower~circle})
\nonumber\\
&=-\frac{1}{16\pi^2}\int_0^{\infty}\frac{d\alpha}{\alpha^2}e^{-i\alpha m^2-\alpha \epsilon}.
\label{FV24}
\end{align}
As we see, $D(x^{\mu}=0,{\rm front})$ is dependent on $m$ after all. We recognize (\ref{FV24}) as (\ref{FV15}), and thus by direct evaluation confirm that the instant time and light front vacuum bubbles are equal, with both being nonzero.

 Now, instead of having to deal with pole or circle contributions, we can also evaluate $D(x^{\mu}=0,{\rm front})$ by using the exponential regulator directly on the real $p_+$ axis. We do this below to obtain $D(x^{\mu}=0,{\rm front},{\rm regulator})$ as given in (\ref{FV30}) below, and recognize (\ref{FV30}) as being none other than (\ref{FV24}). We thus confirm the validity of (\ref{FV24}).

That we were able to avoid pole terms altogether in deriving (\ref{FV24}) is because we used different complex $p_+$ plane contours for $p_->0$ (upper half $p_+$ plane) and $p_-<0$ (lower half $p_+$ plane). However to make contact with the Fock space evaluation we must restrict the discussion to just the one contour that is closed below the real $p_+$ axis. Then, since we can set $D(x^{\mu}=0,{\rm front})=D(x^{\mu}=0,{\rm front},{\rm pole})+D(x^{\mu}=0,{\rm front},{\rm lower~circle})$, we see that, even without evaluating it explicitly,  not only must the circle contribution be nonvanishing, it must restore the dependence on $m$. Thus the correct determination of $D(x^{\mu}=0,{\rm front})$ is its $m$-dependent value given in (\ref{FV24}), and this determination is nonzero.

\section{Reconciling the Fock Space and Feynman Calculations}

Now while we have seen that $D(x^{\mu}=0,{\rm front})$ is not given by $D(x^{\mu}=0,{\rm front},{\rm Fock})$, this is nonetheless puzzling since on the face of it there would not appear to be anything wrong with the Fock space calculation that we have presented above. However, the quantity $-i\langle \Omega|\phi(0)\phi(0)]|\Omega\rangle$ involves the product of two fields at the same spacetime point and is thus ill-defined. To define it we must first split the points and then carefully monitor the limit in which the point splitting is set to zero. We thus use $D(x^{\mu}\neq 0, {\rm front})$ as a point-splitting regulator. Since we have seen that there are issues with both  the $p_-=0$ region and the circle at infinity, we shall avoid them both by evaluating $D(x^{\mu}\neq 0, {\rm front})$ directly on the real $p_+$ axis via the exponential regulator technique, and then monitor the limit in which we set $x^{\mu}$ to zero. This will enable us to develop a single formalism in which we can realize both Fock and Feynman prescriptions simultaneously, so that we can then see exactly where the Fock space approach breaks down. 

In order to do this we will need to represent time-ordering theta functions in a form that also does not involve closing a  Feynman contour. We shall thus employ the real frequency axis exponential regulator for the theta function as well, and set 
\begin{eqnarray}
\theta(x^+)&=&\frac{1}{2\pi }\int_{-\infty}^{\infty}  d\omega \int_0^{\infty}d\alpha e^{-i\omega x^+}e^{i\alpha(\omega+i\epsilon)}
\nonumber\\
&=&\int_0^{\infty}d\alpha e^{-\alpha \epsilon}\delta(\alpha-x^+)
=\int_0^{\infty}d\alpha e^{-x^+ \epsilon}\delta(\alpha-x^+)=\int_0^{\infty}d\alpha \delta(\alpha-x^+),
\nonumber\\
\label{FV25}
\end{eqnarray}
with the $i\epsilon$ providing convergence at $\alpha=\infty$. That $\int_0^{\infty}d\alpha \delta(\alpha-x^+)$ indeed is $\theta(x^+)$ follows since $\int_0^{\infty}d\alpha \delta(\alpha-x^+)=1$ if $x^+$ is positive, and $\int_0^{\infty}d\alpha \delta(\alpha-x^+)=0$ if $x^+$ is negative.

To  evaluate $D(x^{\mu},{\rm front},{\rm regulator})$ with $x^{\mu}\neq 0$ and with no restriction on the sign of  $x^+$, we  set

\begin{eqnarray}
&&D(x^{\mu},{\rm front},{\rm regulator})
\nonumber\\
&&=-\frac{2i}{(2\pi)^4}\int_{-\infty}^{\infty}dp_+\int_{-\infty}^{\infty} dp_1\int_{-\infty}^{\infty} dp_2\int_{-\infty}^{\infty} dp_-e^{-i(p_+x^++p_-x^-+p_1x^1+p_2x^2)}
\int_0^{\infty}
d\alpha e^{i\alpha(4p_+p_--(p_1)^2-(p_2)^2-m^2+i\epsilon)}
\nonumber\\
&&=-\frac{2i}{(2\pi)^3}\int_{-\infty}^{\infty} dp_1\int_{-\infty}^{\infty} dp_2\int _{0}^{\infty}dp_-e^{-i(p_-x^-+p_1x^1+p_2x^2)}
\int_0^{\infty}d\alpha e^{i\alpha(-(p_1)^2-(p_2)^2-m^2+i\epsilon)}\delta(4\alpha p_--x^+)
\nonumber\\
&&-\frac{2i}{(2\pi)^3}\int_{-\infty}^{\infty} dp_1\int_{-\infty}^{\infty} dp_2\int _{-\infty}^{0}dp_-e^{-i(p_-x^-+p_1x^1+p_2x^2)}
\int_0^{\infty}d\alpha e^{i\alpha(-(p_1)^2-(p_2)^2-m^2+i\epsilon)}\delta(4\alpha p_--x^+),
\label{FV26}
\end{eqnarray}
with suppression of the $\alpha$ integration at $\alpha=\infty$ again being supplied by the $i\epsilon$ term. 
On changing the signs of $p_-$, $p_1$ and $p_2$ in the last integral and  setting $F^2_p$ equal to the positive $(p_1)^2+(p_2)^2+m^2$ we obtain
\begin{eqnarray}
&&D(x^{\mu},{\rm front},{\rm regulator})
\nonumber\\
&&=-\frac{2i}{(2\pi)^3}\int_{-\infty}^{\infty} dp_1\int_{-\infty}^{\infty}dp_2\int _{0}^{\infty}\frac{dp_-}{4p_-}e^{-i(p_-x^-+p_1x^1+p_2x^2)}
\int_0^{\infty}d\alpha e^{ix^+(-F^2_p+i\epsilon)/4p_-}\delta(\alpha -x^+/4p_-)
\nonumber\\
&&-\frac{2i}{(2\pi)^3}\int_{-\infty}^{\infty} dp_1\int_{-\infty}^{\infty}dp_2\int _{0}^{\infty}\frac{dp_-}{4p_-}e^{i(p_-x^-+p_1x^1+p_2x^2)}
\int_0^{\infty}d\alpha e^{ix^+(F^2_p-i\epsilon)/4p_-}\delta(\alpha +x^+/4p_-).
\label{FV27}
\end{eqnarray}
Then, using (\ref{FV25}), and with the sign of $p_-$ not being negative  we obtain
\begin{eqnarray}
&&D(x^{\mu},{\rm front},{\rm regulator})
\nonumber\\
&&=-\frac{2i\theta(x^+)}{(2\pi)^3}\int_{-\infty}^{\infty} dp_1\int_{-\infty}^{\infty}dp_2\int _{0}^{\infty}\frac{dp_-}{4p_-}e^{-i(F^2_px^+/4p_-+p_-x^-+p_1x^1+p_2x^2+ix^+\epsilon/4p_-)}
\nonumber\\
&&-\frac{2i\theta(-x^+)}{(2\pi)^3}\int_{-\infty}^{\infty} dp_1\int_{-\infty}^{\infty}dp_2\int _{0}^{\infty}\frac{dp_-}{4p_-}e^{i(F^2_px^+/4p_-+p_-x^-+p_1x^1+p_2x^2-ix^+\epsilon/4p_-)},
\label{FV28}
\end{eqnarray}
and note that the structure of (\ref{FV28}) is such that for $x^+>0$ (forward in time) one only has positive energy propagation, while for $x^+<0$ (backward in time) one only has negative energy propagation. With the insertion into $D(x^{\mu})=-i\langle \Omega|[\theta(x^+)\phi(x)\phi(0)+\theta(-x^+)\phi(0)\phi(x)]|\Omega\rangle$ of the Fock space expansion for $\phi(x^{\mu})$ given in (\ref{FV14}) precisely leading to (\ref{FV28}), we recognize  (\ref{FV28}) as the $x^{\mu}\neq 0$ $D(x^{\mu},{\rm front},{\rm Fock})$ \cite{footnote4}.

Now if we set $x^{\mu}=0$ in (\ref{FV28}) we would appear to obtain the $m$-independent $D(x^{\mu}=0,{\rm front},{\rm Fock})$ given in (\ref{FV19}). However, we cannot take the $x^+\rightarrow 0$ limit since the quantity $x^+/4p_-$ is undefined if $p_-$ is zero, and $p_-=0$ is included in the integration range. Hence, just as discussed in regard to (\ref{FV11}),  the limit is singular. 

To obtain a limit that is not singular we note that we can set $x^{\mu}$ to zero in (\ref{FV26}) as there the limit is well-defined, and this leads to 
\begin{eqnarray}
&&D(x^{\mu}=0,{\rm front},{\rm regulator})
\nonumber\\
&&=-\frac{2i}{(2\pi)^3}\int_{-\infty}^{\infty} dp_1\int_{-\infty}^{\infty}dp_2\int _{0}^{\infty}dp_-
\int_0^{\infty}d\alpha e^{i\alpha(-(p_1)^2-(p_2)^2-m^2+i\epsilon)}\delta(4\alpha p_-)
\nonumber\\
&&-\frac{2i}{(2\pi)^3}\int_{-\infty}^{\infty} dp_1\int_{-\infty}^{\infty}dp_2\int _{-\infty}^{0}dp_-
\int_0^{\infty}d\alpha e^{i\alpha(-(p_1)^2-(p_2)^2-m^2+i\epsilon)}\delta(4\alpha p_-)
\nonumber\\
&&=-\frac{2i}{(2\pi)^3}\int_{-\infty}^{\infty} dp_1\int_{-\infty}^{\infty}dp_2\int _{-\infty}^{\infty}dp_-
\int_0^{\infty}\frac{d\alpha}{4\alpha} e^{i\alpha(-(p_1)^2-(p_2)^2-m^2+i\epsilon)}\delta(p_-)
\label{FV29}
\end{eqnarray}
If we do the momentum integrations we obtain the $m$-dependent
\begin{eqnarray}
&&D(x^{\mu}=0,{\rm front},{\rm regulator})=-\frac{1}{16\pi^2}\int_0^{\infty}\frac{d\alpha}{\alpha^2}e^{-i\alpha m^2-\alpha\epsilon}.
\label{FV30}
\end{eqnarray}
We recognize (\ref{FV30}) as being of the same form as the $m$-dependent $D(x^{\mu}=0,p_->0,{\rm front},{\rm upper~circle})+D(x^{\mu}=0,p_-<0,{\rm front},{\rm lower~circle})$ given in (\ref{FV24}). We thus have to conclude that the limit $x^{\mu}\rightarrow 0$ of (\ref{FV28}) is not (\ref{FV19}) but is (\ref{FV30}) instead. The technical difference between (\ref{FV30}) and (\ref{FV28}) is that to obtain (\ref{FV28}) we did the $\alpha$ integration first, while to obtain (\ref{FV30}) we did the $p_-$ integration first. Only the latter procedure takes care of the $p_-=0$ contribution. Thus to conclude, we see that because of singularities we first have to point split, and when we do so we find that it is the $m$-dependent (\ref{FV30}) that is the correct value for the light front vacuum graph.

\section{Infinite Momentum Frame Considerations}
\label{S16}

The infinite momentum frame is a very convenient frame to use in quantum field theory since many Feynman diagrams are suppressed if an observer makes a Lorentz boost with a velocity at or close to the velocity of light.  Under a Lorentz boost with velocity $u$ in the 3-direction the contravariant and covariant components of a general four-vector $A^{\mu}$ transform as 
\begin{eqnarray}
A^0\rightarrow\frac{A^0+uA^3}{(1-u^2)^{1/2}}, \quad A^3\rightarrow \frac{A^3+uA^0}{(1-u^2)^{1/2}},\quad
A_0\rightarrow\frac{A_0-uA_3}{(1-u^2)^{1/2}}, \quad A_3\rightarrow \frac{A_3-uA_0}{(1-u^2)^{1/2}}.
\label{FV31}
\end{eqnarray}
If we set $(1-u)=\epsilon^2/2$, then with $\epsilon$ small, to leading order we  obtain
\begin{eqnarray}
&&A^0\rightarrow\frac{A^0+A^3}{\epsilon} +O(\epsilon), \quad A^3\rightarrow \frac{A^3+A^0}{\epsilon}+O(\epsilon),\quad
A_0\rightarrow\frac{A_0-A_3}{\epsilon}+O(\epsilon), \quad A_3\rightarrow \frac{A_3-A_0}{\epsilon}+O(\epsilon),
\nonumber\\
&&(A^0)^2-(A^3)^2\rightarrow A^+A^-+O(\epsilon).
\label{FV32}
\end{eqnarray}
This leads to
\begin{eqnarray}
p^3\rightarrow  \frac{p^+}{\epsilon}=\frac{2p_-}{\epsilon}, \quad
E_p\rightarrow \frac{2p_-}{\epsilon}+\frac{((p_1)^2+(p_2)^2+m^2)\epsilon}{4p_-}+....,\quad 
\frac{dp^3}{E_p}\rightarrow \frac{dp_-}{p_-},
\label{FV33}
\end{eqnarray}
where $E_p=((p_3)^2+(p_1)^2+(p_2)^2+m^2)^{1/2}$.

As well as transform energies and momenta we also have to transform the ranges of integration in Feynman graphs. To this end we recall that under a Lorentz boost the velocity transforms as
\begin{eqnarray}
v \rightarrow \frac{v+u}{1+vu}=\frac{v+1-\epsilon^2/2}{1+v-v\epsilon^2/2}.
\label{FV34}
\end{eqnarray}
Thus with $u=1-\epsilon^2/2$, $v=1-\epsilon^2/2$ transforms into $v^{\prime}=1$, while $v=-1+\epsilon^2/2$ transforms into $v^{\prime}= -1$. With the quantity $p^3+p^0$ being given by $m(v+1)/(1-v^2)^{1/2}=m(1+v)^{1/2}(1-v)^{1/2}$, the range $p^3=-\infty$ to $p^3=+\infty$, viz. $v=-1+\epsilon^2/2$ to $v=1-\epsilon^2/2$, transforms into the $p^+=2p_-$ range $m\epsilon^2/4$ to $\infty$, and thus to the range $0$ to $\infty$ when we set $\epsilon=0$.

In the instant time vacuum sector we had found that  $D(x^{\mu}=0,{\rm instant},{\rm Fock})$ and $D(x^{\mu}=0,{\rm instant},{\rm pole})$ are equal, with both being given by (\ref{FV15}). On transforming (\ref{FV15}) to the infinite momentum frame and comparing with (\ref{FV17}) and (\ref{FV19}) we obtain 
\begin{eqnarray} 
&&D(x^{\mu}=0,{\rm instant},{\rm Fock})=D(x^{\mu}=0,{\rm instant},{\rm pole})= -\frac{i}{(2\pi)^3}\int_{-\infty}^{\infty} \frac{d^3p}{2E_p}
\nonumber\\
&&\rightarrow -\frac{i}{(2\pi)^3}\int _{-\infty}^{\infty} dp_1\int _{-\infty}^{\infty}dp_2\int _0^{\infty}\frac{dp_-}{2p_-}
=D(x^{\mu}=0,{\rm front},{\rm Fock})=D(x^{\mu}=0,{\rm front},{\rm pole}).
\label{FV35}
\end{eqnarray}
As such, the infinite momentum frame is doing what it is supposed to do, namely it is transforming an instant time on-shell graph into a light front on-shell graph. However, we have seen that the light front mass-independent on-shell evaluation of the vacuum graph does not agree with correct mass-dependent value provided by the off-shell light front vacuum Feynman diagram.  Thus in this respect not only is the on-shell prescription failing for light front vacuum graphs, so is the infinite momentum frame prescription. (Technically, one ordinarily applies the infinite momentum frame approach to instant time Feynman diagrams, but as long as they only receive pole contributions this is equivalent to applying the infinite momentum frame approach to the instant time on-shell Fock space amplitude.)

There is an oddity in (\ref{FV35}), one peculiar to the infinite momentum frame. Since the mass-dependent quantity $d^3p/2E_p$ is Lorentz invariant, under a Lorentz transformation with a velocity less than the velocity of light it must transform into itself and thus must remain mass dependent. However, in the infinite momentum frame it transforms into a quantity $dp_1dp_2dp_-/2p_-$ that is mass independent. This is because velocity less than the velocity of light and velocity equal to the velocity of light are inequivalent, since an observer that is able to travel at less than the velocity of light is not able to travel at the velocity of light. Lorentz transformations at the velocity of light are different than those at less than the velocity of light, and at the velocity of light observers (viz. observers on the light cone) can lose any trace of mass. 

Moreover, (\ref{FV35}) also raises a puzzle. Specifically,  while the instant time on-shell evaluation of the vacuum $D(x^{\mu}=0,{\rm instant},{\rm Fock})$ does coincide with the instant time evaluation of the vacuum off-shell Feynman diagram  $D(x^{\mu}=0,{\rm instant},{\rm regulator})$ \cite{footnote5}, and while the instant time evaluation of the off-shell Feynman diagram  $D(x^{\mu}=0,{\rm instant},{\rm regulator})$ does coincide with the light front evaluation of the off-shell Feynman diagram  $D(x^{\mu}=0,{\rm front},{\rm regulator})$, nonetheless, the light front on-shell evaluation of the vacuum $D(x^{\mu}=0,{\rm front},{\rm Fock})$ does not coincide with the light front evaluation of the off-shell Feynman diagram $D(x^{\mu}=0,{\rm front},{\rm regulator})$.

The resolution of this puzzle lies in the contribution of circle at infinity to the Feynman contour. In the instant time case the integral $\int dp_0dp_3/[(p_0)^2-(p_3)^2-(p_1)^2-(p_2)^2-m^2+i\epsilon]$ is suppressed on the circle at infinity in the complex $p_0$ plane ($p_3$ being finite), and only poles contribute. However, when one goes to the infinite momentum frame the instant time $dp_3$ also becomes infinite ($p^3=mv/(1-v^2)^{1/2}$) and the circle contribution is no longer suppressed. Specifically,  on the instant time circle at infinity the term that is of relevance behaves as $\int Rie^{i\theta}d\theta dp_3/(R^2e^{2i\theta}-(p_3)^2)$, and on setting $\epsilon=1/R$ in the infinite momentum frame limit it behaves as the nonvanishing $\int Rie^{i\theta}d\theta Rdp_-/(R^2e^{2i\theta}-R^2p_-^2)$.Thus in the instant time case one cannot ignore the circle at infinity in the infinite momentum frame even though one can ignore it for observers moving with finite momentum, a point that appears to have been missed in prior infinite momentum frame studies. Consequently, the initial reduction from the instant time Feynman diagram to the on-shell instant time Fock space prescription is not valid in the infinite momentum frame, and one has to do the full four-dimensional Feynman contour integral.

We had noted earlier a general rule that the on-shell evaluation always coincides with the pole term evaluation, and that if the pole is not the only contributor to the Feynman contour then the Feynman and Fock space prescriptions cannot agree and one must use the Feynman prescription. We can now add that if we ignore the effect of an infinite Lorentz boost on the instant time circle at infinity, the instant time infinite momentum frame evaluation always coincides with the light front pole term evaluation, and if the light front pole is not the only contributor to the light front Feynman contour then the Feynman and infinite momentum frame evaluations cannot agree and one must use the light front Feynman contour or exponential regulator prescription. 

\section{Dressing the Vacuum Graph}

\begin{figure}[H]
\begin{center}
\includegraphics[scale=0.15]{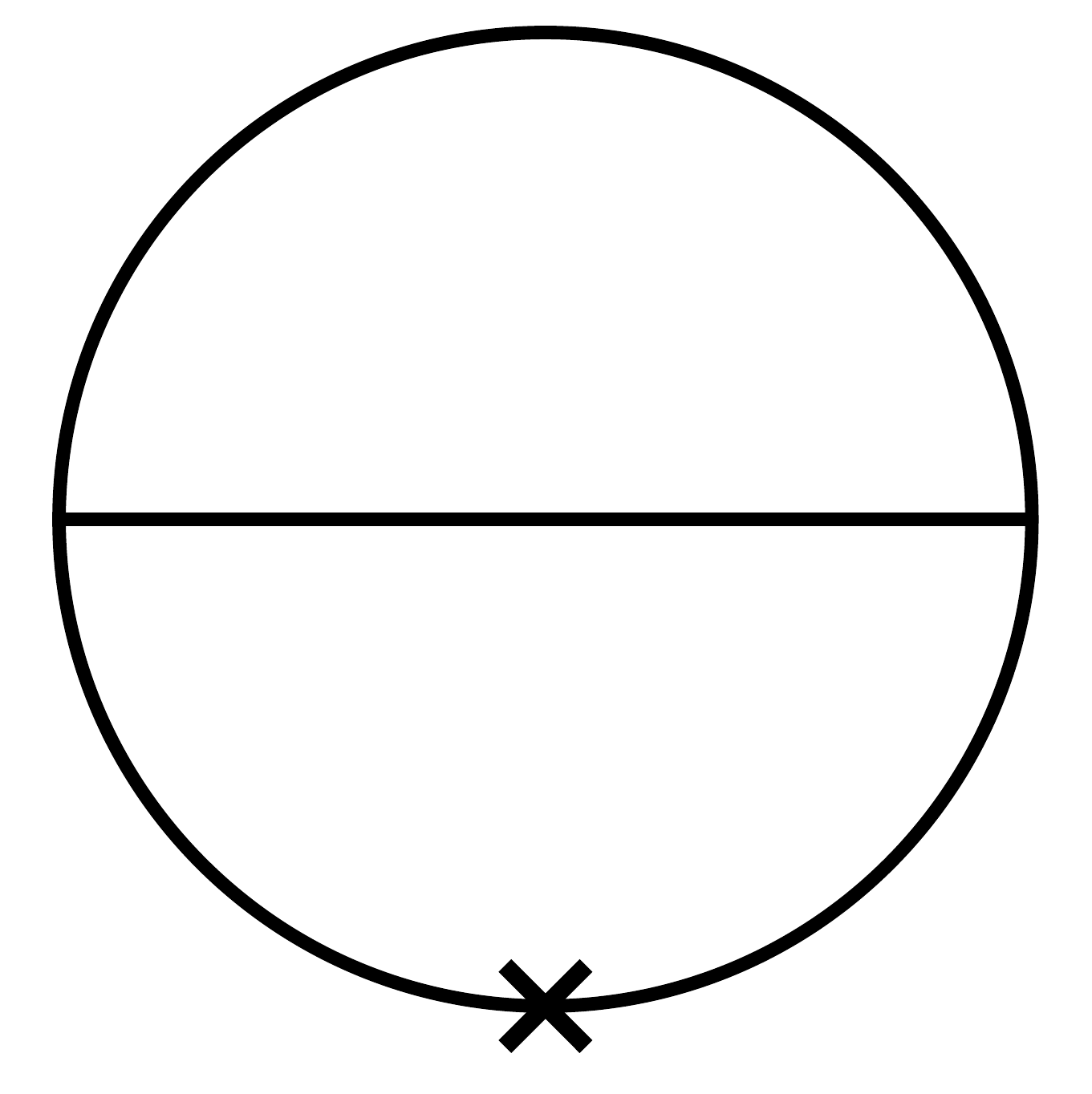}
\caption{Connected dressed vacuum graph}
\label{vacuumtadpoledressed}
\end{center}
\end{figure}

The first dressing to Fig. \ref {vacuumtadpole} is shown in Fig. \ref{vacuumtadpoledressed}. This graph is actually a self-energy graph within a vacuum loop. To see this, momentarily separate the lines at the cross. This is then a $\Sigma(p)$ self-energy renormalization graph. However, this renormalization comes with a $\delta m$ and a $Z$, and the graph can be replaced by a dressed propagator. To calculate 
\begin{eqnarray}
I=\int \frac{d^4kd^4p}{(4p_+p_--F^2_p+i\epsilon)(4k_+k_--F^2_k+i\epsilon)[4(k_++p_+)(k_-+p_-)-F^2_{p+k}+i\epsilon]}
\label{FV36}
\end{eqnarray}
(viz. Fig. \ref{vacuumtadpoledressed}), we first do the $d^4k$ integration with $p_{\mu}$ held fixed. Up to irrelevant factors this is
\begin{eqnarray}
\Sigma(p)=\int \frac{dk_+dk_1dk_2dk_-}{(4k_+k_--k_1^2-k_2^2-m^2+i\epsilon)[4(k_++p_+)(k_-+p_-)-(k_1+p_1)^2-(k_2+p_2)^2-m^2+i\epsilon]}.
\label{FV37}
\end{eqnarray}
There is no circle at infinity contribution as the denominator has two powers of $k_+$. The graph diverges as a single logarithm, i.e. as ${\rm log}(\Lambda^2/p^2)$ at large $p^2$. Introducing a mass renormalization counter term $\delta m=-{\rm log}(\Lambda^2/m^2)$ gives  ${\rm log}(m^2/p^2)$. The original graph is thus 
\begin{eqnarray}
I=\int dp_+dp_1dp_2dp_-\frac{{\rm log}(m^2/p^2)}{(4p_+p_--p_1^2-p_2^2-m^2+i\epsilon)}
\label{FV38}
\end{eqnarray}
and the circle at infinity is not suppressed. The concerns raised in this paper thus carry over to dressed light front vacuum graphs as well and cannot be ignored.

\section{Shortcomings of Normal Ordering}

In our study of light front vacuum graphs we have studied a point of principle, namely the appropriate way to evaluate the vacuum graphs. Now a reader might regard the issue as being somewhat  academic since vacuum graphs can be normal ordered away. However, since (perturbative) normal ordering involves moving all annihilation operators to the right and all creation operators to the left in a vacuum matrix element, it does not encompass the circle at infinity contributions that occur in light front vacuum graphs. For light front vacuum graphs we thus need to deal with circle at infinity contributions and such contributions are foreign to standard renormalization techniques, and indeed in their presence one cannot effect a Wick rotation to Euclidean momenta. To get round this we note that for renormalization one does not actually need to consider circle at infinity contributions per se since one can evaluate Feynman diagrams as real frequency integrals by using the exponential regulator on the real frequency axis, just as was done for the light front case in (\ref{FV26}). Then one can introduce a second field with a regulator mass $M$ and subtract off its contribution, the Pauli-Villars prescription, and use the Pauli-Villars prescription to regulate the ultraviolet behavior of light front vacuum graphs.

Moreover, we also note that certain vacuum graphs are actually observable and cannot in fact be normal ordered away anyway, namely  those associated with dynamical symmetry breaking or those that couple to gravity. When a symmetry is broken dynamically by a fermion bilinear condensate one is interested in evaluating the expectation value $\langle S|\bar{\psi}\psi|S\rangle$ where $|S\rangle$ is a spontaneously broken vacuum and then comparing it with the expectation value $\langle N|\bar{\psi}\psi|N\rangle$ where $|N\rangle$ is a normal vacuum. Now dynamical symmetry breaking is a long range order infrared effect while normal ordering or Pauli-Villars is a way of dealing with ultraviolet divergences. Since dynamical symmetry breaking is an infrared effect the short distance behaviors of $\langle S|\bar{\psi}\psi|S\rangle$ and $\langle N|\bar{\psi}\psi|N\rangle$ are the same. Thus even if we were to normal order $\langle N|\bar{\psi}\psi|N\rangle$ by setting $\langle N|:\bar{\psi}\psi:|N\rangle$ equal to zero, we would still need to evaluate $\langle S|:\bar{\psi}\psi:|S\rangle$ where the normal ordering is done with respect to $|N\rangle$, with $\langle S|:\bar{\psi}\psi|:S\rangle$ being nonzero in the broken symmetry case.

For gravity, consider a free massive fermion with energy-momentum tensor $T_{\mu\nu}=i\bar{\psi}\gamma_{\mu}\partial_{\nu}\psi$. The Einstein equations are of the form $R_{\mu\nu}-(1/2)g_{\mu\nu}R=-8\pi G T_{\mu\nu}$ (here $R$ is the Ricci scalar), with trace $R=8\pi G m\langle \Omega|\bar{\psi}\psi|\Omega \rangle$. The fermion vacuum bubble thus couples to gravity. Moreover, here one is not free to normal order at all, since the hallmark of Einstein gravity is that gravity couples to energy and not to energy difference. Since in analog to $\langle\Omega|\phi^2|\Omega \rangle$ the light front circle at infinity contribution to $\langle \Omega|\bar{\psi}\psi|\Omega \rangle$ is nonzero, in the light front the circle at infinity contributes to the cosmological constant. 

\section{Conclusions}

If one starts with a field equation and an equal instant time or equal light front time commutator one can make a Fock space expansion of a field in terms of creation and annihilation operators as multiplied by on-shell solutions to the field equation. (For an instant time free field for instance the solutions are labelled by a three-vector $\bar{p}$ with the energy fixed to the on-shell $E_p=(p^2+m^2)^{1/2}$). In constructing a Feynman diagram one has the same field equations and the same canonical commutators but one in addition has a time ordering. It is this time ordering that takes the Feynman diagram off shell, with the energy being replaced by a contour integration in a complex frequency plane as in (\ref{FV4}). In the literature there are four approaches to dealing with the light front vacuum sector: the Feynman diagram approach, the light front Fock space approach, the light front Fock space approach as restricted to states with $p_->0$, and the light front sector as derived by writing the instant time vacuum sector in the infinite momentum frame.  In this paper we have analyzed all of these different approaches and identified why they differ and identified which approach (viz. the Feynman diagram approach) is to be the valid one.

In the non-vacuum sector all of these approaches lead to the same outcome because in the Feynman diagrams there are only pole contributions (to thus recover both the Fock space and infinite momentum frame approaches), and when $x^+$ is non-zero and positive (the scattering situation) the $\delta(4\alpha p_- -x^+)$ term in (\ref{FV26}) ensures that only $p_->0$ terms are relevant. However, in the light front vacuum sector there are circle at infinity contributions in the Feynman diagrams to thus make the Fock space and infinite momentum frame approaches incorrect, while the replacement of the $\delta(4\alpha p_- -x^+)$ term by $\delta(4\alpha p_-)$ when $x^+$ is zero now permitting a contribution from $p_-=0$. Thus as noted in going from (\ref{FV4}) with $x^+\neq 0$ to (\ref{FV5}) with $x^+=0$ one has to include a circle at infinity contribution that one previously had not needed to include. It is this circle at infinity contribution that is then paramount  in the light front vacuum sector, to thus make the off-shell Feynman diagram approach with its non-zero value for light front vacuum graphs the correct one.

\begin{acknowledgments}
PDM would like to thank Dr. T. G. Rizzo for the kind hospitality of the Theory Group at the Stanford Linear Accelerator Center where part of this work was performed. The work of SJB and the work of PL were supported in part by the Department of Energy under contract DE-AC02-76SF00515.
\end{acknowledgments}


\begin{thebibliography}{99}

\bibitem{Dirac:1949cp}\href{https://doi.org/10.1103/RevModPhys.21.392}{ P. A. M. Dirac,
  Rev.  Mod.  Phys. \textbf {21}, 392 (1949).}
  
 \bibitem{Brodsky:1997de}\href{https://doi.org/10.1016/S0370-1573(97)00089-6}{S. Brodsky, H.-C. Pauli and  S. Pinsky, Phys. Rept. \textbf{301}, 299 (1998).}
 
\bibitem{Leutwyler:1977vy}\href{https://doi.org/10.1016/0003-4916(78)90082-9}{H. Leutwyler and J. Stern,
  Annals Phys. \textbf{112}, 94 (1978).}
 
  

\bibitem{Bakker:2013cea} \href{https://doi.org/10.1016/j.nuclphysbps.2014.05.004}{B. L. G. Bakker {\it et al.},
  Nucl. Phys. B Proc.  Suppl.   \textbf{251-252}, 165 (2014).}
 

\bibitem{Burkardt:1995ct}\href{https://doi.org/10.1007/0-306-47067-5_1}{M.~Burkardt,
  Adv. Nucl. Phys.  \textbf{23}, 1 (1966).}
   
\bibitem{Lepage:1980fj}\href{https://doi.org/10.1103/PhysRevD.22.2157}{G. P. Lepage and S. J. Brodsky,
  Phys. Rev. D \textbf{22}, 2157 (1980).}
   
\bibitem{Lepage:1979zb}\href{https://doi.org/10.1016/0370-2693(79)90554-9}{G. P. Lepage and S. J. Brodsky,
  Phys. Lett. B \textbf{87}, 359 (1979).}
    

\bibitem{Brodsky:2003pw}\href{https://doi.org/10.1103/PhysRevD.69.076001}{S. J. Brodsky, J. R. Hiller, D. S. Hwang and V. A. Karmanov, Phys. Rev. D \textbf{69}, 076001 (2004).}
  
 \bibitem{Weinberg:1966jm}\href{https://doi.org/doi:10.1103/PhysRev.150.1313}{S.Weinberg, Phys. Rev. \textbf{ 150}, 1313 (1966).}

  
 
\bibitem{Brodsky:1973kb}\href{https://doi.org/10.1103/PhysRevD.8.4574}{S. J. Brodsky, R. Roskies and R. Suaya, Phys. Rev. D \text{8}, 4574 (1973).}
  
  







\bibitem{Chang1969} \href{https://doi.org/10.1103/PhysRev.180.1506}{S.-J. Chang and S. K. Ma, Phys. Rev. \textbf{180}, 1506 (1969).}

\bibitem{Yan1973} \href{https://doi.org/10.1103/PhysRevD.7.1780}{T.-M. Yan, Phys. Rev. D \textbf{7}, 1780 (1973).}

\bibitem{Casher1974} \href{https://doi.org/10.1103/PhysRevD.9.436}{A. Casher and L. Susskind, Phys. Rev. \textbf{D} 9, 436 (1974).} 

\bibitem{Brodsky:2009zd}\href{https://doi.org/10.1073/pnas.1010113107}{S. J. Brodsky and R. Shrock, Proc. Nat.  Acad. Sci.   \textbf{108}, 45 (2011).}



\bibitem{Collins2018}\href{https://arxiv.org/abs/1801.03960}{J. Collins. \textit{The non-triviality of the vacuum in light-front quantization: An elementary treatment}, arXiv:1801.03960 [hep-ph], January 2018.}


\bibitem{footnote1} In light front coordinates the line element is given by $ds^2=x^+x^--(x^1)^2-(x^2)^2$, to thus correspond to a metric with $g_{+-}=g_{-+}=1/2$, $g_{11}=g_{22}=-1$, ${\rm det}[g_{\mu\nu}]=-1/4$, and $x_+=x^-/2$, $x_-=x^+/2$. The invariant measure is given by $(1/2)\int dx^+dx^1dx^2dx^-=2\int dx_+dx_1dx_2dx_-$.

\bibitem{Mannheim2019} P. D. Mannheim, P. Lowdon and S. J. Brodsky, {\it Comparing light front quantization with instant time quantization}, SLAC-PUB-17390, February, 2019.

\bibitem{footnote1a} Equations obeyed by spacetime-dependent scalar field Green's functions such as $[g^{-1/2}\partial_{\nu}g^{1/2}\partial^{\nu}+m^2]D(x^{\mu})=-g^{-1/2}\delta ^4(x)$ are general coordinate invariant c-number equations, and one can directly transform them from instant time coordinates to light front coordinates. An equivalent argument is made in \cite{Mannheim2019} for the the scalar field c-number path integral formulation, and this enables us to extend the general coordinate invariance argument to the spacetime-independent vacuum sector.

\bibitem{Neville1971} \href{https://doi.org/10.1007/BF02734389}{R. A. Neville and F. Rohrlich,  Il Nuovo Cimento A \textbf{1}, 625 (1971).}

\bibitem{footnote2} In passing, we note that it is often found advantageous in  the light front parton literature to restrict the Fock space expansion to hypersurfaces with $x^+=0$. Here we make no such restriction.

\bibitem{footnote3} In the spontaneously broken case it had been noted in \href{https://doi.org/10.1103/PhysRevD.66.045019}{P. P. Srivastava and S. J. Brodsky Phys. Rev. D \textbf{66}, 045019 (2002)} that there then is a $p_+=0$, $p_-=0$, $p_1=0$, $p_2=0$ soft mode contribution.

\bibitem{footnote4} While the on-shell old-fashioned Hamiltonian perturbation theory formalism actually predates quantum field theory, as  (\ref{FV28}) shows, even with quantum field theory it continues to hold in the light front case for forward in time scattering processes with $x^+>0$, since for them the energy $F^2_p/4p_-$ is positive. However, the on-shell approach fails for the light front vacuum graphs that appear in quantum field theory, with the $\theta(-x^+)$ term making a  contribution in the $x^+\rightarrow 0$ limit, with the $x^+\rightarrow 0$ limit being singular, and with the intuition acquired for $x^+>0$ not carrying over to the light front $x^+=0$ vacuum sector. From the perspective of the on-shell light front $4p_+p_--(p_1)^2-(p_2)^2=m^2$ condition, $p_-=0$ corresponds to $p_+=\infty$, and thus to $x^+=0$, so that  the $p_-\rightarrow 0$  and the $x^+\rightarrow 0$ limits are both consequential. In contrast, we should note that in the instant time case the $x^0\rightarrow 0$ limit is not singular, with the structure of the vacuum sector in the light front case thus being conceptually different from its structure in the instant time case. 

\bibitem{footnote5} In \cite{Mannheim2019} the instant time $D(x^{\mu}=0,{\rm instant},{\rm regulator})$  is evaluated using the exponential regulator on the real $p_0$ axis. Its value is found to be equal to $D(x^{\mu}=0,{\rm instant})$ as given in (\ref{FV15}), just as it should be.


\end{thebibliography}
\end{document}